\documentclass[a4paper,12pt]{article}
\usepackage{amsmath}
\usepackage{amssymb}
\usepackage{amsfonts}
\usepackage{graphics}
\usepackage{epsfig}

\begin{document}
\renewcommand{\theequation}{\thesection.\arabic{equation}}
\thispagestyle{empty}
\vspace*{-1.5cm}

\begin{center}
{\Large\bf { Local cubic vertex functions for three massless higher even spin fields on spaces $AdS_{D}$: An analytic approach}}\\
\vspace{4cm}
\large{Werner R\"uhl}\\
Department of Physics, Technical University of Kaiserslautern\\P.O.Box 3049,
67653 Kaiserslautern, Germany \\
\end{center}
\vspace{6cm}

\begin{abstract}
Local cubic vertex functions of three higher even spin fields on $AdS_{D}$ are constructed from the Green function of three conserved currents that are dual to the higher spin fields. Conservation of the currents implies lowest order gauge invariance.
These vertex functions appear by the $UV$ divergence as the residue of the highest order pole in the dimensional regularization parameter $\epsilon$. In fact $N$-point Green functions of such conserved currents produce a series of poles up to the order $N-1$. The method works for even $D$ and maintains covariance at any step. The resulting formula is quite concise.
\end{abstract}
\vspace{3cm}
\begin{center}
April 2013
\end{center}

\newpage

\section{Introduction}
\setcounter{equation}{0}

It is the aim of this article to construct vertex functions for three massless higher spin fields of even spin on $AdS$ spaces. This construction is done in such fashion that these vertex functions are local and observe gauge invariance of lowest order.
The history of cubic vertex functions on flat or (anti)deSitter spaces is long and has put forward different aspects of interest. First we must mention the seminal work of Fradkin and Vasiliev \cite{FV}. The frame-like approach to vertex functions of any order and non-abelian gauge symmetry proposed by these authors has developed in recent decades into several directions (for a recent source see \cite{BPS}).

In flat spaces the general 3-vertex functions were constructed in \cite{MMR1} and restrictions on the form of these vertex function that were achieved earlier \cite{M} could be verified. Connection with string theory was emphasized in \cite{ST}. Contrary to the very satisfactory results for flat spaces, the situation for $AdS$ spaces is still
not completely satisfactory at least in the sense that the resulting formulae do not exhibit such a simple  and beautiful form as in the flat case. There are different methods applied to solve the $AdS$ cases. Using flat ambient spaces and then reducing the dimension by one looks as a convincing ansatz but technical complications show up when this is done explicitly \cite{JT}, \cite{MMR2}. Instead of this ansatz we use now a method based on quantum field theory and the regularization techniques of UV divergences leading to a compact final formula.

We contract a higher spin field of a symmetric tensor representation of rank ("spin") $s$ with a conserved current of the same representation. This current is built bilinearly from a real scalar conformal free field admitting only even rank tensors. The vacuum expectation value of three such currents is UV divergent, and this divergence
can be characterized by a polynomial in $\epsilon^{-1}$, where this parameter $\epsilon$ can be introduced by a deformation of the dimension $D$ of the $AdS$ space. For $N$ such currents the $N$-point Green function yields a UV divergence described by a polynomial of degree $N-1$. We are interested here only in the residue of the highest order pole term. It is known to preserve a symmetry such as gauge invariance.
This method has been developed for flat spaces \cite{WR}, in which case we were also able to successfully compare the results with the general known formulas \cite{MMR1}. It seems that the postulate of $AdS$ covariance introduces new problems. The reader will find, however, that the choice of an appropriate mathematical
formalism simplifies the whole task dramatically. This formalism was developed in \cite{MR1} a few years ago when the quantum one loop trace anomaly of the same fields and currents was analysed for $AdS_4$. The restriction to the dimension $D=4$ was motivated by the desire to compare the results with
known anomalies in gravitation theory. In the present context such motivation does not seem to exist and we can as well and will deal with cases of any even space
dimension. The relevant formulae of \cite{MR1} are generalized for this purpose to any even dimension $D$.

All traces of the higher spin fields are neglected, so that our
result holds for traceless higher spin fields. This can be justified
by two arguments: The physically relevant terms are still present in
the result, and the trace terms are uniquely determined by the
traceless parts. The $AdS$ space radius $L$ is, as is often done,
fixed to the value $1$. An arbitrary value $L$ can be recovered by
replacing the square bracket of $D_k$ defined in (\ref{4.14}) and in
the final formula (\ref{5.1}) by $\Box +L^{-2}[\Delta(\Delta+1)
-n(n+1)]$ . An essential part of the formalism developed in this
article uses a general basis of bitensors \cite{AT} to express the
two-point functions of the currents. The same technique was
developed to present two-point functions of higher spin fields in
\cite{LMR} and applied to two-point functions of currents in
\cite{MR1}.

\section{Scalar fields and conserved currents}
\setcounter{equation}{0} It is remarkable though of course natural
that we go back to the analysis of anomalies of 2-loop functions on
$AdS_{D}$ in \cite{MR1} when we study the singular part of N-loop
functions. The scalar field and the currents are in fact the same.
The free scalar conformal field  $\sigma(z)$ has the two point
function \cite{CSHF}
\begin{equation}
<\sigma(z_1)\quad\sigma(z_2)> = w(\zeta),\quad  \zeta-1 = \frac{(z_1-z_2)^2}{2z_{1,0} z_{2,0}} = u
\label{2.1}
\end{equation}
where Poincare coordinates are used. The function $w(\zeta)$ is a
Legendre function of the second kind with the desired asymptotic
behaviour which in terms of a Gaussian hypergeometric function is
\begin{equation}
w(\zeta) = \frac{\Gamma(\Delta)}{(2\pi)^{\frac{D}{2}}}\,\zeta^{-\Delta}\,F(\frac{\Delta}{2},\frac{\Delta+1}{2};
\frac{1}{2};\zeta^{-2})
\label{2.2}
\end{equation}
and the parameter $\Delta$ denotes
\begin{equation}
\Delta = \frac{D}{2}-1
\label{2.3}
\end{equation}
where $\Delta$ is the conformal dimension of the scalar field $\sigma$.
For arbitrary $D$ the two-point function (\ref{2.2}) is identical with the binomial expansion of
\begin{equation}
w(\zeta) = \frac{\Gamma(\Delta)}{2(2\pi)^{\frac{D}{2}}}\,[ \frac{1}{(\zeta-1)^{\Delta}} + \frac{1}{(\zeta+1)^{\Delta}}]
\label{2.4}
\end{equation}
that is rational for even $D$. For odd $\Delta (D=4n)$
replacing $\zeta$ by $-\zeta$ changes the sign of the function $w(\zeta)$. This symmetry is denoted $antipodal$ $symmetry$, the sign change $antipodal$ $parity$. We recognize
that the 2-point function has two singular points at $\zeta = +1$ and $\zeta = -1$, such pair is called an antipodal pair.  Correspondingly a 3-point function of three currents has eight maximal singular points when
\begin{equation}
\zeta_{1,2} = \pm 1,\, \zeta_{2,3} = \pm 1,\, \zeta_{3,1} = \pm 1
\label{2.5}
\end{equation}
Analogously the 3-vertex using (2.4) consists of eight parts. We
shall choose that one which belongs to
\begin{equation}
\zeta_{1,2} = \zeta_{2,3} = \zeta _{3,1} = 1
\label{2.6}
\end{equation}
and concentrate our work on this alone.

For the conserved current we can take the expression $J^{(s)}(z;a)$
from \cite{MR1}, equ. (1) or from \cite{Ans}, $s$ is even. The
complete expression of $J^{(s)}(z;a)$ is a polynomial in $a^2$ of
degree $\frac{s}{2}$ and each factor of $(a^2)^{n}$ is a polynomial
in $L^{-2}$ ($L$ is the radius of the $AdS$ space) of degree $n$.
Since we want to consider only the traceless terms of the cubic
vertex, we concentrate only on the $(a^2)^0$ part
\begin{equation}
J^{(s)}(z;a) = 1/2\sum_{p=0}^{s} A_{p}^{(s)}
(a\nabla)^{s-p}\sigma(z) (a\nabla)^{p}\sigma(z) + \textnormal{$a^2$
terms} \label{2.7}
\end{equation}
 The coefficients are
\begin{equation}
A_{p}^{(s)} = (-1)^{p}{s \choose p}\frac{(D/2 -2)!(s+D/2-2)!}{(p+D/2-2)!(s-p+D/2-2)!}
\label{2.8}
\end{equation}
We are interested in the loop Green function
\begin{equation}
<J^{(s_1)}(z_1;a_1) \quad J^{(s_2)}(z_2;a_2) \quad J^{(s_3)}(z_3;a_3)>
\label{2.9}
\end{equation}
This Green function is evaluated by Wick's theorem using
(\ref{2.1}).

\section{Evaluating the Green function}
\setcounter{equation}{0}

Let $F(\zeta)$ be an analytic function of $\zeta$. Then we can use
the formulae (24) and (25) of \cite{MR1},
\begin{eqnarray}
&&(a,\nabla_1)^{p} (b,\nabla_2)^{q} F(\zeta(z_1,z_2)) =
\sum_{n=0}^{q-p} \frac{p!q!}{n!(q-n)!(p-q+n)!}\nonumber\\ &&\times
I_{1,2}^{q-n}I_1(1,2)^{p-q+n} I_2(1,2)^{n} \,
F^{(p+q)}(\zeta(z_1,z_2)) + a^2,b^2\,\textnormal{terms}\label{3.1}
\end{eqnarray}
using the bitensor basis

\begin{eqnarray}
I_1(1,2) = (a,\partial_1)\zeta(z_1,z_2) \qquad\qquad\\
I_2(1,2) = (b,\partial_2)\zeta(z_1,z_2) \qquad\qquad\\
I_{1,2} =(a,\partial_1)(b,\partial_2)\zeta(z_1,z_2)\qquad\qquad
\label{3.4}
\end{eqnarray}
where
\begin{equation}
F^{(k)}(\zeta) = \frac{d^{k}}{d\zeta^{k}}\, F(\zeta)
\label{3.5}
\end{equation}
For the Green function we get in this fashion (up to trace terms)
\begin{eqnarray}
\sum_{p_1= 0}^{s_1}\sum_{p_2=0}^{s_2}\sum_{p_3=0}^{s_3} \, A_{p_1}^{(s_1)} A_{p_2}^{(s_2)} A_{p_3}^{(s_3)} \,
(a,\nabla_1)^{p_1}(b,\nabla_2)^{s_2-p_2}\,w(\zeta_{1,2}) \nonumber\\
\times (b,\nabla_2)^{p_2}(c,\nabla_3)^{s_3-p_3}\,w(\zeta_{2,3})\,
(c,\nabla_3)^{p_3}(a,\nabla_1)^{s_1-p_1}\,w(\zeta_{3,1})
\label{3.6}
\end{eqnarray}
which by equs. (3.1) to (3.4) yields
\begin{eqnarray}
\sum_{p_1=0}^{s_1}\sum_{p_2=0}^{s_2}\sum_{p_3=0}^{s_3} A_{p_1}^{(s_1)}\,A_{p_2}^{(s_2)}\,A_{p_3}^{(s_3)} \textbf{Q}
_{p_1,p_2,p_3}^{(s_1,s_2,s_3)} \qquad \nonumber\\ \times\quad w^{(p_1+s_2-p_2)}(\zeta_{1,2})\,w^{(p_2+s_3-p_3)}
(\zeta_{2,3}) \,w^{(p_3+s_1-p_1)}(\zeta_{3,1})
\label{3.7}
\end{eqnarray}
where the newly introduced function $\bf{Q}$ depends besides the parameters $s_{i},p_{i}$ on the three points $z_1,z_2,z_3$.
It encodes the tensorial structure of the 3-point vertex function. The number of derivatives of the functions $w$ in (\ref{3.7})
are denoted by
\begin{equation}
m_{i,i+1} = p_{i} + s_{i+1} - p_{i+1}
\label{3.8}
\end{equation}
A simple formula for $\textbf{Q}$ is
\begin{eqnarray}
 \textbf{Q} _{p_1,p_2,p_3}^{(s_1,s_2,s_3)} = \sum_{n_{1,2}=max\{0,s_2-p_2-p_1\}}^{s_2-p_2}\quad\sum_{n_{2,3}=max\{0,s_3-p_3-p_2\}}^{s_3-p_3} \quad\sum_{n_{3,1}=max\{0, s_1-p_1-p_2\}}^{s_1-p_1}
\nonumber\\ (\Delta)_{m_{12}}(\Delta)_{m_{23}}(\Delta)_{m_{31}}
\frac{(s_1-p_1)!(s_2-p_2)!(s_3-p_3)!}{n_{1,2}!\,n_{2,3}!\,n_{3,1}!}\qquad\qquad\nonumber\\ \times
 {p_1 \choose s_2-p_2-n_{1,2}}{p_2 \choose s_3-p_3-n_{2,3}}{p_3 \choose s_1-p_1-n_{3,1}}\qquad\qquad\nonumber\\
\times [I_1(1,2)I_2(1,2)/I_{1,2}]^{n_{1,2}}I_1(1,2)^{p_1+p_2-s_2}\, [I_2(2,3)I_3(2,3)/I_{2,3}]^{n_{2,3}} I_2(2,3)^{p_2+p_3-s_3}\nonumber\\ \times[I_1(3,1)I_3(3,1)/I_{3,1}]^{n_{3,1}}I_3(3,1)^{p_3+p_1-s_1}
I_{1,2}^{s_2-p_2} I_{2,3}^{s_3-p_3} I_{3,1}^{s_1-p_1}\qquad\qquad
\label{3.9}
\end{eqnarray}
Obviously the sums over the parameters $n_{1,2}, n_{2,3}, n_{3,1}$ can be performed in terms of hypergeometric $_1F_1 $
polynomials. The result has a cyclic order
\begin{eqnarray}
\textbf{Q}_{p_1,p_2,p_3}^{(s_1,s_2,s_3)} =(\Delta)_{m_{12}}(\Delta)_{m_{23}}(\Delta)_{m_{31}}\nonumber\\
\times \frac{(s_2-p_2)!}{(s_2-p_1-p_2)!}\,I_2(1,2)^{s_2-p_1-p_2}I_{1,2}^{p_1}\nonumber\\
\times _1F_1(-p_1;s_2-p_1-p_2+1; -\frac{I_1(1,2)I_2(1,2)}{I_{1,2}})\nonumber\\
\times \frac{(s_3-p_3)!}{(s_3-p_2-p_3)!}\,I_3(2,3)^{s_3-p_2-p_3}I_{2,3}^{p_2} \nonumber\\
\times _1F_1(-p_2;s_3-p_2-p_3+1;-\frac{I_2(2,3)I_3(2,3)}{I_{2,3}})\nonumber\\
\times \frac{(s_1-p_1)!}{(s_1-p_1-p_3)!}\,I_1(3,1)^{s_1-p_3-p_1}I_{3,1}^{p_3} \nonumber\\
\times _1F_1(-p_3;s_1-p_3-p_1+1;-\frac{I_3(3,1)I_1(3,1)}{I_{3,1}})
\label{3.10}
\end{eqnarray}
If, say, $s_2-p_2-p_1$ is negative, in the first factor the function $_1F_1$ starts at the term
$ [I_1(1,2)I_2(1,2)]^{p_1+p_2-s_2}$, thus replacing essentially the factor $I_2(1,2)^{s_2-p_1-p_2}$ in front of $_1F_1$ by $I_1(1,2)^{p_1+p_2-s_2}$. In either expression a zero at $z_1-z_2$ of order $\mid s_2-p_1-p_2\mid$ is contained. A closer look at the zeros of $\textbf{Q}$ will be presented in Section 5.

\section{The regularization of the $w$-functions}
\setcounter{equation}{0} Remember the definition of the UV divergent
part of (\ref{2.1}) with default normalization
\begin{eqnarray}
w(\zeta) = (\zeta-1)^{-\Delta},  \quad \Delta = \frac{D}{2} -1 \\
u = \zeta -1 \\
w^{(n)}(\zeta) = (-1)^{n}(\Delta)_{n}(\zeta-1)^{-\Delta-n}
\label{4.3}
\end{eqnarray}
where $D$ is even and $(\Delta)_{n}$ is a Pochhammer symbol.
Moreover we restrict $D$ to $\geq 4$. We intend to use the method of
"dimensional regularization" by introducing a parameter $\epsilon$
interpreted as a deformation of the dimension $D$ of the $AdS_{D}$
space. Then the regularized 3-vertex function appears as a rational
function of $\epsilon$ with a pole of maximal order 2 (for an
$N$-vertex loop function it would be $N-1$). We select this pole
from our vertex function (3.7) - (3.10) since it delivers us the
local differential operator defining the interaction Lagrangian
density.

According to \cite{MR1} equ. (47) we have
\begin{equation}
\frac{1}{u^{\Delta+n-\epsilon}} = \frac{(-1)^{\Delta+n-1}}{\epsilon (\Delta+n-1)!} \delta^{(\Delta+n-1)}(u) + O(1)
\label{4.4}
\end{equation}
where $\epsilon$ is thought of being hidden in $\Delta = D/2-1$ as a deformation of $D$. This justifies the term "dimensional" regularization.
Moreover we use \cite{MR1} equ. (75) ($\Omega_{D-1}$ is the area of the unitsphere in $D$ dimensions)
\begin{equation}
d\mu(z) = (2z_0)^{-D} d^{D}z = [u(u+2)]^{\Delta}\, du\, d\Omega_{D-1}, \quad u = \frac{(z_1-z_2)^2}{2z_1^0\,z_2^0}   \label{4.6}
\end{equation}
where the polar coordinates in $AdS_{D}$ are defined by $z_2=z$ and
$z_1$ as the pole (reference point)
\begin{eqnarray}
(2z_1^0)^{D} \delta(z_2-z_1) =
\frac{(-1)^{\Delta}\delta^{(\Delta)}(u)}{\Delta!
(u+2)^{\Delta}\Omega_{D-1}} \qquad\qquad \label{4.6}
\end{eqnarray}

We must treat the distribution ($w$ with the normalization from
(4.1))
\begin{eqnarray}
lim_{\epsilon\rightarrow 0+}\quad\epsilon^2 w^{(m_{12})} (\zeta_{12}) w^{(m_{23})}
(\zeta_{23}) w^{(m_{32})} (\zeta_{31})
\nonumber \\
= (-1)^{m_{12}+m_{23}+m_{31}}(\Delta)_{m_{12}}(\Delta)_{m_{23}}(\Delta)_{m_{31}} \nonumber\\
\times lim_{\epsilon\rightarrow 0+}\quad\epsilon^2 u_{12}^{-m_{12}-\Delta +\epsilon} u_{23}^{-m_{23}-\Delta +\epsilon}
u_{31}^{-m_{31}-\Delta +\epsilon}
\label{4.7}
\end{eqnarray}
This distribution resulting in the limit is determined only by the geometry of the $AdS$ space. We treat it
in coordinate space. In flat space Fourier transforms are used and momenta are integrated over. Instead of
applying corresponding harmonic analysis on $AdS$, we will be content, however, with presenting the structure
of the distribution and not all the explicit algebraic expressions.
Then define
\begin{equation}
r_{12}= \Delta + m_{12} -1
\label{4.8}
\end{equation}
so that (4.4) goes into
\begin{equation}
u_{12}^{-r_{12}- 1 +\epsilon} = \frac{1}{\epsilon}
\frac{(-1)^{r_{12}}}{r_{12}!}\delta^{(r_{12})}(u_{12}) \label{4.9}
\end{equation}
We conclude from (4.9) that a deltafunction of $u$ has the minimal
derivative $\Delta$ in order to define a distribution on $AdS_{D}$
space. Moreover (4.9) shows that negative $r_{12}$ do not
contribute.

In order to express deltafunctions with argument $u_{12}$ by
deltafunctions with argument $z_2 - z_1$ we use a method developed
in \cite{MR1}. We start from
\begin{equation}
\Phi_{n}(u) = \frac{\delta^{(n)}(u)}{(u+2)^{\Delta}}
\label{4.10}
\end{equation}
and apply the differential operator $(u+2)^{\Delta} \Box$ (with the scalar Laplacian)
\begin{eqnarray}
(u+2)^{\Delta} \Box \, \Phi_{n}(u) = A_{n} \delta^{(n+1)}(u)  +B_{n} \delta^{(n)}(u) \\
A_{n} = 2(\Delta -n-1), \quad B_{n} = n(n+1) -\Delta(\Delta+1)
\label{4.12}
\end{eqnarray}
This allows us to formulate the recursion \footnote{In order to
recover the $L$ dependence we replace in (\ref{4.14}) $\Box
+\Delta(\Delta+1)-n(n+1)$ by $\Box+L^{-2}[\Delta(\Delta+1)
-n(n+1)]$}
\begin{eqnarray}
\Phi_{n+1}(u) = -D_{n}\Phi_{n}(u) \qquad \qquad\\
D_{n} = \frac{1}{2(n+2)-D} \,[\Box +\Delta(\Delta+1) - n(n+1)]
\label{4.14}
\end{eqnarray}
In the case of the variables $u_{12}$ or $u_{31}$ we place the pole
of the coordinate system at $z_1$ , define corresponding
differential operators $D_{n}$ and $\Box$ acting on these
coordinates ($n=2$ respectively $n=3$) and denote them
correspondingly by $D_{n}(2), \Box(2)$ (respectively $D_{n}(3),
\Box(3)$). Then we get e.g. by solving the recursion (4.13), (4.14)
and starting from $\Phi_{\Delta}$ using (4.6)
\begin{equation}
\Phi_{n+\Delta}(u_{12}) = (-1)^{n+\Delta}\Delta! \{\prod_{k=\Delta}^{n+\Delta-1} D_{k}(2)\}
(2z_1^0)^{D} \delta(z_2-z_1) \Omega_{D-1}
\label{4.15}
\end{equation}

Now we return from $\Phi_{n}$ to the delta functions
\begin{eqnarray}
\delta^{(n+\Delta)}(u_{12})&=& (u_{12}+2)^{\Delta} \Phi_{n+\Delta}(u_{12}) \nonumber \\
&=& \sum_{\ell = 0}^{\Delta} \frac{(-1)^{\ell}\Delta!\,2^{\Delta-\ell}(n+\Delta)! }{\ell!(\Delta-\ell)!
(n+\Delta-\ell)!} \Phi_{n+\Delta-\ell}\,(u_{12}) \label{4.16}\\
&=& (\Delta!)^2\sum_{\ell=0}^{\Delta} \frac{(-1)^{n+\Delta}2^{\Delta-\ell}
(n+\Delta)!}{\ell!(\Delta-\ell)!
(n+\Delta-\ell)!} \nonumber\\
&& \times \{\prod_{k=\Delta}^{n+\Delta-\ell-1} D_{k}(2)\}
(2z_1^0)^{D}\delta(z_2-z_1) \Omega_{D-1} \label{4.17}
\end{eqnarray}
and we consider the case of $\zeta = \zeta_{1,2}$. In this case we
have to replace  $n+\Delta$ by $r_{12} = m_{12} + \Delta -1$ and to
multiply with (see (4.8), (4.9))
\begin{equation}
\frac{(-1)^{r_{12}}}{r_{12 }!}
\label{4.18}
\end{equation}
to obtain the distribution part of $w^{(m_{12})}(\zeta_{12})$. We
introduce then a shorthand for the differential operator (4.17)
\begin{equation}
\Lambda_{r_{12}}(2) = \sum_{\ell\geq 0}\frac{2^{\Delta-\ell}}{\ell!(\Delta-\ell)!(r_{12}-\ell)!}
\{\prod_{k_1=\Delta}^{r_{12}-\ell} D_{k_1}(2)\}
\label{4.19}
\end{equation}
that acts on the deltafunction
\begin{equation}
(2z_1^0)^{D}\delta(z_2-z_1)
\label{4.20}
\end{equation}

Now we integrate partially with the result (where $\ast_{a}$ denotes
contraction over $a$)
\begin{equation}
(2z_1^0)^{D} \delta(z_2-z_1) \Lambda_{r_{12}}(2)\{u_{23}^{-m_{23}-\Delta} u_{31}^{-m_{31}-\Delta} \times
\textbf{Q} \ast_{a_1} \ast_{a_2} \ast_{a_3}h^{(s_1)} h^{(s_2)} h^{(s_3)}(z_1,z_2,z_3) \}
\label{4.21}
\end{equation}
and denote from now on the factors behind the multi-cross by the
shorthand
\begin{equation}
\textbf{R}_{p_1,p_2,p_3}(z_1,z_2,z_3) = \textbf{Q} \ast_{a_1} \ast_{a_2} \ast_{a_3} h^{(s_1)} h^{(s_2)} h^{(s_3)}
(z_1,z_2,z_3)
\label{4.22}
\end{equation}
First we study the case that $\Lambda_{r_{12}}$ acts on
$\textbf{R}$. Then there results, replacing $u_{23}$ by $u_{31}$
using the deltafunction $\delta(z_2-z_1)$
\begin{equation}
(2z_1^0)^{D} \delta (z_2-z_1) u_{31}^{-2\Delta-m_{23}-m_{31}} \times \Lambda_{r_{12}} \textbf{R}
\label{4.23}
\end{equation}
This power of $u_{31}$ is denoted $-r_{31}-1$ and deformed by adding
$2\epsilon$. Then the leading UV divergent term is
\begin{equation}
\frac{1}{2\epsilon} \frac{(-1)^{r_{13}}}{r_{31}!} \delta^{(r_{31})}(u_{31})
\label{4.24}
\end{equation}
and we can proceed as before and define a new differential operator
\begin{eqnarray}
\Lambda_{r_{31}}(3)  = \sum_{\ell\geq 0} \frac{2^{\Delta-\ell}}{l!(\Delta-\ell)!(r_{31}-\ell)!} \{\prod_{k_2=\Delta}^{r_{31}-\ell-1} D_{k_2}(3)\}
\label{4.25}
\end{eqnarray}
so that we end up in this case after partial integration with
\begin{equation}
(2z_1^0)^{D}\delta(z_2-z_1) (2z_1^0)^{D} \delta(z_3-z_1) \Lambda_{r_{12}}(2) \Lambda_{r_{31}}(3) \textbf{R}(z_1,z_2,z_3)
\label{4.26}
\end{equation}

However there is the more general case that
\begin{eqnarray}
\Lambda_{r_{12}}(2) \{ u_{23}^{-m_{23}-\Delta} \textbf{R} \}
= \sum_{\kappa} \Theta_{\kappa}^{(I)} u^{-m_{23} -\Delta}
\Theta_{\kappa}^{(II)} \textbf{R}
\label{4.27}
\end{eqnarray}
by Leibniz's rule, where the case that for a special $\kappa_0$ we
have
\begin{equation}\label{4.28}
\Theta_{\kappa_0}^{(II)} = \textnormal{const}
\end{equation}
has already been dealt with.

Consider the expression
\begin{equation}
\Theta_{\kappa}^{(I)} u_{23}^{-m_{23}-\Delta}
\label{4.29}
\end{equation}
In the resulting sum we consider the terms with the same factor
\begin{equation}
u_{23}^{-m_{23} -\Delta -q}, q \geq 0
\label{4.30}
\end{equation}
Then we get
\begin{equation}
\Theta_{\kappa}^{(I)} u_{23}^{-m_{23}-\Delta} = \sum_{q \geq 0} \textbf{P}_{\kappa,q}(z_2-z_3, z_2^0, z_3^0)
u_{23}^{-m_{23}-\Delta-q}
\label{4.31}
\end{equation}
where $\textbf{P}$ is a polynomial in $z_2-z_3$, but squares
$(z_2-z_3)^2$ are excluded. This negative power of $u_{23}$ has to
be regularized as before in (4.19) and leads to the differential
operator $\Lambda_{m_{23}+\Delta+q-1}(2)$.

Thus after the usual partial integration we end up with the
expression
\begin{eqnarray}
(2z_1^0)^{D}\delta (z_2-z_1) (2z_1^0)^{D}\delta(z_3-z_1) \sum_{q} \Lambda_{m_{23} +\Delta +q -1}(2)
\nonumber\\ \sum_{\kappa} \textbf{P}_{\kappa,q}(z_2-z_1, z_2^0,z_1^0) \Theta_{\kappa}^{(II)}(2)
\Lambda_{m_{31} +\Delta -1}(3)\textbf{R}_{p_1,p_2,p_3}
\label{4.32}
\end{eqnarray}

\section{Discussion}
\setcounter{equation}{0}

By partial integration the polynomials of the Laplacians $\Box(2), \Box(3)$ (see (\ref{4.20})) acting on delta functions
can be brought to act on the product of $\textbf{Q}$ and the higher spin fields. The result is
\begin{eqnarray}
\int \frac{d^{D}z}{(2z^0)^{D}}\,\sum_{p_1=0}^{s_1}\sum_{p_2=0}^{s_2}\sum_{p_3=0}^{s_3}\,A_{p_1}^{(s_1)}A_{p_2}^{(s_2)}A_{p_3}^{(s_3)} (\Delta)_{m_{12}}
(\Delta)_{m_{23}} (\Delta)_{m_{31}} \nonumber\\ \times
\sum_{q \geq 0}\Lambda_{m_{23} + \Delta +q -1}(2) \textbf{P}_{\kappa,q}(z_2-z_1,z_2^0,z_1^0)\Theta_{\kappa}^{(II)}
(2) \Lambda_{m_{31} +\Delta -1}(3)
\qquad\qquad \nonumber\\ \times
\textbf{Q}_{p_1,p_2,p_3}^{(s_1,s_2,s_3)} *_{a_1}
*_{a_2}
*_{a_3}\quad h^{(s_1)}(z,a_1) h^{(s_2)}(z_2,a_2)
h^{(s_3)}(z_3,a_3)\mid_{z_2=z_3= z} \qquad\qquad \label{5.1}
\end{eqnarray}
where the shorthands (see (\ref{3.8}))
\begin{equation}
m_{i,i+1} = p_{i} +s_{i+1}-p_{i+1}
\label{5.2}
\end{equation}
and (4.25), (4.27), (4.31) have been used. The asterisk symbols
denote contractions that produce a scalar function of $\textbf{Q}$
and $h^{(s_1)}h^{(s_2)}h^{(s_3)}$.

The Laplacians  and the gradients (from $\Theta_{\kappa}^{II}(2)$)
act only on the variables $z_2$ and $z_3$. This is the effect of a
partial integration eliminating all differentiations with respect to
$z_1$. Other approaches to these vertex functions may give results
symmetric in the three variables $z_1 ,z_2, z_3$, which makes any
comparison with our result troublesome. In any case differentiations
on the factors contained in $\textbf{Q}$ in (3.9) or (3.10) seem to
necessitate an algorithmic computer program.

The maximal number of differentiations is
\begin{equation}
2(r_{12}+r_{31} -2) = 3D +2S  -4,\quad S = s_1+s_2+s_3
\label{5.3}
\end{equation}
However, in $\textbf{Q}$ there are zeros which have to be cancelled first before the differentiations act on the fields. These zeros are hidden in $I_{i}(i,i+1), I_{i+1}(i,i+1)$ and are each of order one. Thus the total number of zeros in $\textbf{Q}$ is
\begin{equation}
\Psi = \mid s_1-p_1-p_3\mid + \mid s_2-p_2-p_1 \mid + \mid s_3-p_3-p_2 \mid
\label{5.4}
\end{equation}
Now assume that the three numbers $s_1, s_2, s_3$ satisfy triangular inequalities, then we can solve $\Psi = 0$ by
\begin{equation}
p_1 = \frac{1}{2}(s_1+s_2-s_3),\quad p_2 = \frac{1}{2}(s_2+s_3-s_1), \quad p_3 = \frac{1}{2}(s_3+s_1-s_2)
\label{5.5}
\end{equation}
In this case the maximal number of derivatives acting on the fields
is the one given by (5.3).

Acknowledgement: Critical remarks and suggestions by R. Manvelyan are gratefully acknowledged. We thank the Galileo Galilei Institute for 
Theoretical Physics for the hospitality and the INFN for partial support during the completion of this work.

\end{document}